\documentclass[3p,times,sort&compress]{elsarticle}

\usepackage{graphicx}
\usepackage{amsmath,amssymb,color}
\usepackage{epstopdf,pdflscape,float,hyperref}

\newcommand{\ds}{\displaystyle}
\newcommand{\e}{{\rm e}}

\begin{document}

\begin{frontmatter}
\title{Exact time-dependent solutions of a Fisher-KPP-like equation obtained with nonclassical symmetry analysis}
\author[qut]{Scott W McCue \corref{cor1}}
\author[UniSA]{Bronwyn H Bradshaw-Hajek}
\author[qut]{Matthew J Simpson}
\address[qut]{School of Mathematical Sciences, Queensland University of Technology, Brisbane, Australia.}
\address[UniSA]{UniSA STEM, University of South Australia, Mawson Lakes SA 5095, Australia.}
\cortext[cor1]{Corresponding author: scott.mccue@qut.edu.au}

\begin{abstract}
We consider a family of exact solutions to a nonlinear reaction-diffusion model, constructed using nonclassical symmetry analysis.  In a particular limit, the mathematical model approaches the well-known Fisher-KPP model, which means that it is related to various applications including cancer progression, wound healing and ecological invasion. The exact solution is mathematically interesting since exact solutions of the Fisher-KPP model are rare, and often restricted to long-time travelling wave solutions for special values of the travelling wave speed.
\end{abstract}

\end{frontmatter}

\section{Introduction} \label{sec:intro}

Reaction-diffusion models are commonplace in applied mathematics.  These involve one or more parabolic partial differential equations (pdes) that, in one dimension, are written as
\begin{equation}
\frac{\partial u}{\partial t}=\frac{\partial}{\partial x}\left(D(u)\frac{\partial u}{\partial x}\right)+R(u).
\label{eq:generalreactiondiffusion}
\end{equation}
We discuss possible forms for $D(u)$ and $R(u)$ shortly, but note that for the purposes of our study, one of the most popular examples of (\ref{eq:generalreactiondiffusion}) is the Fisher-KPP equation (in dimensionless form)
\begin{equation}
\frac{\partial u}{\partial t}=\frac{\partial^2 u}{\partial x^2}+u(1-u),
\label{eq:FisherKPP}
\end{equation}
\cite{Fisher1937,Kolmogorov1937}, which is used extensively.  The overall theme of our work is to study a family of exact solutions of a version of (\ref{eq:generalreactiondiffusion}) that are closely related to solutions of (\ref{eq:FisherKPP}).

More generally, in reaction-diffusion models of the form (\ref{eq:generalreactiondiffusion}), the nonlinear reaction term $R(u)$ is used to model reproduction in ecology~\cite{Melica2014,Simpson2022} or proliferation in cell biology~\cite{Vittadello2018,Sherratt90}.   A common form for this reaction term is (in dimensional variables) the logistic growth term $R(u)=\lambda u(1-u/K)$, where $\lambda$ is the reproduction rate and $K$ is the carrying capacity.  Indeed, this is the reaction term in (\ref{eq:FisherKPP}).  A feature of the logistic term is that the per capita growth rate $\lambda(1-u/K)$ linearly decreases to zero at $u=K$.  Other qualitatively similar monostable reaction terms $R(u)$ exist that retain key properties including a single local maximum and $R(0)=R(K)=0$~\cite{Jin2016,Simpson2022}.  In this sense, logistic growth falls into a class of qualitatively similar reaction terms.

Forms for nonlinear diffusion $D(u)$ in (\ref{eq:generalreactiondiffusion}) vary depending on the application, although linear diffusion is the most commonly used form (as in the Fisher-KPP model (\ref{eq:FisherKPP})).  One option is to take the nonlinear diffusion to be $D(u)=u^n$, where $n>0$~\cite{Sherratt90,Mccue2019}.  Of particular relevance to our study here, in the context of modelling biological cells, is where $D(u)$ is a decreasing function of $u$~\cite{Cai2007}.

Much attention has been devoted to deriving exact solutions to versions of (\ref{eq:generalreactiondiffusion}).  Travelling wave solutions to the Fisher-KPP equation were first presented by Ablowitz and Zepatella \cite{Ablowitz1979} and then others~\cite{Kaliappan1984,McCue2021}.  Here, a number of travelling wave solutions exist \cite{Kametaka1976,McKean1970,Rinzel1975}. Periodic solutions have been found \cite{Carpenter1977,Hastings1974}, and nonclassical symmetry solutions have been constructed~\cite{Arrigo1994,Clarkson1994} (these solutions can also be found using Painlev\'{e} analysis \cite{Conte1988,Chen1992}). Fewer solutions exist when the diffusivity is nonconstant (and the reaction term is nonzero). Nonclassical symmetry solutions for particular forms of $D(u)$ and $R(u)$ have been presented in Refs~\cite{Arrigo1995,Goard1996,Broadbridge2016,BradshawHajek2019,BradshawHajek2020}.

\begin{figure}
	\centering
	\includegraphics[width=0.60\textwidth]{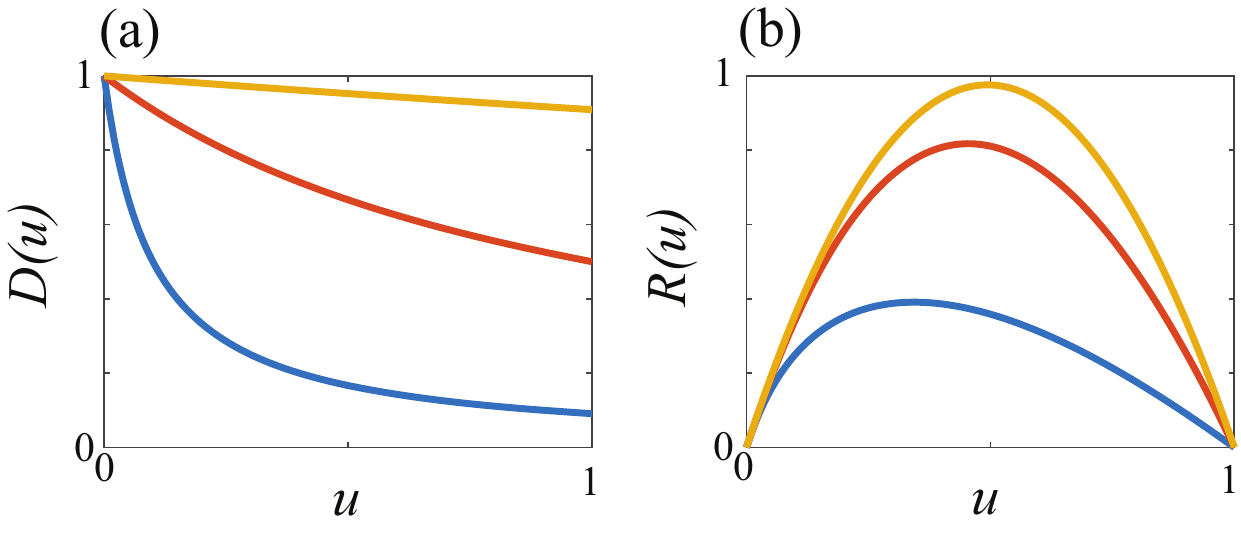}
	\caption{\textbf{Nonlinear diffusion $D(u)$ and source term $R(u)$}. Functions are given for $\beta = 0.1$ (blue), $\beta=1$ (orange) and $\beta=10$ (yellow). }
	\label{fig:figure0}
\end{figure}

Here we report on a family of exact solutions to (\ref{eq:generalreactiondiffusion}),
\begin{equation}
u(x,t)=\beta\left[\mathrm{exp}\left(\ln\left(1+\frac{u_0}{\beta}\right)\mathrm{e}^{-\beta t}\cos \left(\sqrt{1+\beta}\,x\right)\right)-1\right],
\quad 0<x<\frac{\pi}{2\sqrt{1+\beta}},
\label{eq:exact}
\end{equation}
with
\begin{equation}
D(u)=\frac{\beta}{u+\beta},
\quad
R(u)=\beta(1-u)\ln\left(1+\frac{u}{\beta}\right).
\label{eq:DandR}
\end{equation}
The form of $R(u)$ and $D(u)$ in (\ref{eq:DandR}) is such that (\ref{eq:generalreactiondiffusion}) is analogous to (\ref{eq:FisherKPP}), as indicated in Fig.~\ref{fig:figure0}.
Here, $R(u)$ is a monostable reaction term with zeros at $u=0$ and $1$, while $D(u)$ is a positive decreasing function of $u$ for all $u>0$.  In fact, this same precise $D(u)$ has been used to study cell migration~\cite{Cai2007}.  Therefore the exact solutions (\ref{eq:exact}) fall into the same class of solutions to (\ref{eq:FisherKPP}) and we discuss this connection in some detail.

The content of the letter is as follows.  We summarise in Section~\ref{sec:nonclassical} the nonclassical symmetry analysis that leads to (\ref{eq:exact}).  In Section~\ref{oneparameter} we explain how the solutions describe population extinction.  Limits of large and small $\beta$ are treated briefly in Section~\ref{sec:limits}, noting the connection with (\ref{eq:FisherKPP}), while a further illustrative example is provided in Section~\ref{sec:example}.  We close in Section~\ref{sec:discussion} with a discussion.

\section{Derivation and interpretation}\label{sec:derivation}

\subsection{Nonclassical symmetry solution of the nonlinear reaction-diffusion equation}\label{sec:nonclassical}

Classical Lie point symmetry analysis, first introduced by Sophus Lie, provides a systematic way to search for the invariant quantities in a differential equation, where we seek transformations that leave the equation of interest invariant. These classical symmetries can lead to the well-known travelling wave solutions or scale-invariant solutions (for more detail see for example \cite{Olver1982,Ibragimov1994}). Galaktionov, et al. \cite{Galaktionov1988} provided the first complete classical symmetry classification for equations of type \eqref{eq:generalreactiondiffusion}.

The nonclassical symmetry (or Q-conditional) method was first introduced by Bluman and Cole \cite{Bluman1969}, where we again seek transformations that leave the equation of interest invariant, but we also require that the invariant surface condition be satisfied. This can sometimes result in additional symmetries that cannot be found using the classical method. If nonclassical symmetries can be found, they can be used in the same way as those found using the classical method. That is, the differential equation can be simplified and an analytic solution may be constructed. The most complete nonclassical symmetry analysis of equations of type \eqref{eq:generalreactiondiffusion} is given in~\cite{Arrigo1995,Goard1996}.

To construct \eqref{eq:exact} we use of a  nonclassical symmetry admitted by equation \eqref{eq:generalreactiondiffusion} whenever $R(u)$ and $D(u)$ are related by \cite{Arrigo1995,Goard1996}
\begin{equation}
R(u)=\left(\alpha-\ds\frac{\beta}{D(u)}\right)\int_{u^*}^u\,D(u')\, \textrm{d}u',
\label{eq:relationship}
\end{equation}
where $u^*$ is one of the zeros of $R(u)$ (here we choose $u^*=0$) and $\alpha$ and $\beta$ are constants. Equation \eqref{eq:generalreactiondiffusion} can be reduced to the Helmholtz equation $\Psi_{xx}+\alpha\Psi=0$, where $\Psi(x)$ is related to $u(x,t)$ by
\begin{equation}
\ds\int_{u^*}^u\,D(u')\,\textrm{d}u'=\e^{-\beta t}\Psi.
\label{eq:transformation}
\end{equation}
When $\alpha=k^2>0$, solutions to the Helmholtz equation can be written in terms of sine and cosine functions,
\begin{equation}
\Psi(x)=c_1\sin kx+c_2\cos kx,
\label{eq:Psi}
\end{equation}
where $c_1$ and $c_2$ are constants.

A pair of nonlinear reaction and diffusion terms that satisfy relationship \eqref{eq:relationship} and are analogous to the Fisher-KPP equation are
\begin{equation}
D(u)=\ds\frac{\beta(1-b)}{\alpha(u-b)},\qquad R(u)=\beta(1-u)\ln\left(1-\frac{u}{b}\right).
\label{eq:diffusionreaction}
\end{equation}
The solution to equation \eqref{eq:generalreactiondiffusion} with $D(u)$ and $R(u)$ given by \eqref{eq:diffusionreaction} can be found by rearranging transformation \eqref{eq:transformation} to obtain
\begin{equation}
u(x,t)=b\left[1-\exp\left(\ds\frac{\alpha}{\beta(b-1)}\e^{-\beta t}\Psi(x)\right)\right].
\label{eq:nonclasssolution}
\end{equation}
\subsection{One-parameter family of solutions}\label{oneparameter}

To reduce the number of parameters in (\ref{eq:Psi}) and (\ref{eq:nonclasssolution}), we are motivated by the linear problem~\cite{Skellam1951}
\begin{equation}
\frac{\partial u}{\partial t}=\frac{\partial^2 u}{\partial x^2}+u,
\quad 0<x<L,
\label{eq:linear1}
\end{equation}
\begin{equation}
\frac{\partial u}{\partial x}=0 \quad\mbox{on}\quad x=0,
\label{eq:linear2}
\end{equation}
\begin{equation}
u=0, \quad\mbox{on}\quad x=L,
\label{eq:linear3}
\end{equation}
\begin{equation}
u(x,0)=f(x), \quad 0<x<L,
\label{eq:linear4}
\end{equation}
for which there is an exact solution found using separation of variables.  For large time, this exact solution behaves like $u\sim A_1\mathrm{e}^{-(\pi^2/4L^2-1)t}\cos\left(\pi x/2L\right)$, where $A_1$ is related to the initial condition.  In the derivation of (\ref{eq:linear1})-(\ref{eq:linear4}) from a dimensional system, the length-scale and time-scale of the original physical problem are related to the dimensional diffusion coefficient and growth rate parameter in the usual way so that the pde (\ref{eq:linear1}) does not contain any parameters.  Apart from the initial condition (\ref{eq:linear4}), the only parameter in (\ref{eq:linear1})-(\ref{eq:linear4}) is $L$.  The final key point to note about (\ref{eq:linear1})-(\ref{eq:linear4}) is that, because the eigenvalue is $-\pi^2/4L^2+1$, clearly the solution of (\ref{eq:linear1})-(\ref{eq:linear4}) grows without bound if $L>\pi/2$ while it decays (goes extinct) if $L<\pi/2$.

Returning to (\ref{eq:generalreactiondiffusion}), the dynamics of the solution becomes much clearer if we choose the diffusion and reaction functions so that $D(u)\rightarrow 1$ and $R(u)\sim u$ as $u\rightarrow 0$.  In other words, for small population density $u$, we want our nonlinear system to behave close to the linear system (\ref{eq:linear1})-(\ref{eq:linear4}).  As a result, we choose $b=-\beta$, $\alpha=1+\beta$.  Further, to match the boundary conditions (\ref{eq:linear2})-(\ref{eq:linear3}), we choose $c_1=0$.  Finally, to ensure $u(0,0)=u_0$, say, we choose $c_2=-\beta\ln\left(1+u_0/\beta\right)$.

Before moving on, it is worth noting we are free to replace $t$ with $t-t_0$ in (\ref{eq:exact}), since the original pde (\ref{eq:generalreactiondiffusion}) is invariant under translations in time.  However, the term $\ln(1+u_0/\beta)\mathrm{e}^{\beta t_0}$ is a constant, so changing $t_0$ is equivalent to redefining $u_0$.  Therefore it is only worth keeping one of $u_0$ or $t_0$.  We choose to keep the former.

\begin{figure}
	\centering
	\includegraphics[width=0.90\textwidth]{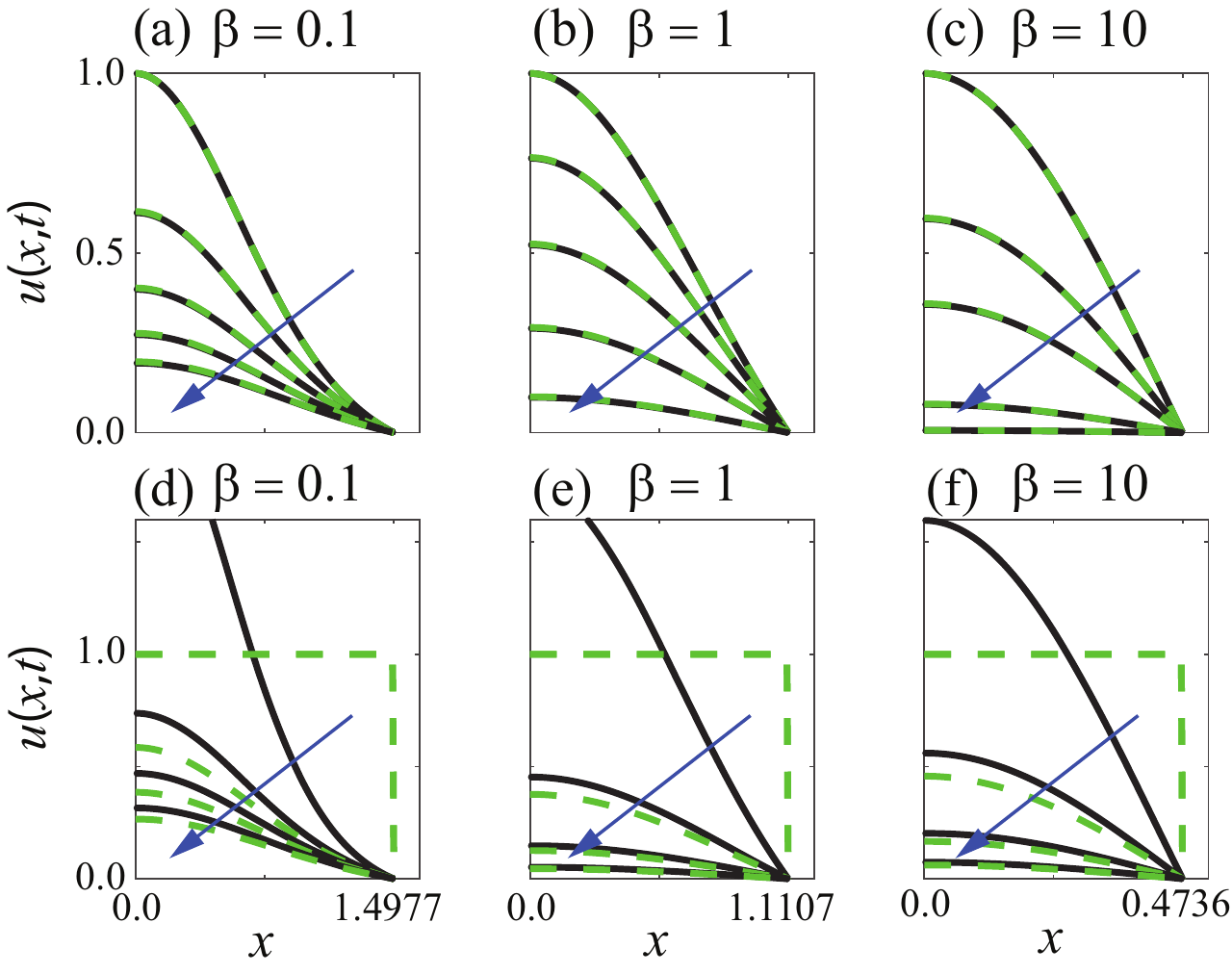}
	\caption{\textbf{Exact time-dependent solutions of Equation (\ref{eq:generalreactiondiffusion})}. (a)-(c) Exact (solid black) and numerical (dashed green) solutions for $\beta=0.1, 1$ and $10$, respectively, with the arrows showing the direction of increasing $t$. Profiles in (a)-(c) are given at $\beta t = 0, 2, 4, 6$ and $8$, $\beta t = 0, 0.2, 0.5, 1$ and $2$, and $\beta t = 0, 1, 2, 5$ and $10$, respectively.  Profiles in (d)-(f) compare numerical solutions for $u(x,0)=1$ with the exact solutions with $u_0$ chosen so that the initial mass is the same in each case and we see that the late-time solutions match reasonably well.  In (d)-(f) profiles are compared at  $\beta t = 0, 4, 6$ and $8$, $\beta t = 0, 1, 2$ and $3$, and $\beta t = 0, 0.1, 0.2$ and $0.3$. Numerical solutions are obtained using the method of lines where the $x$ variable is discretised on a uniform mesh with 201 grid points.}
	\label{fig:figure1}
\end{figure}

With a single parameter $\beta$, we can illustrate the exact solutions with representative values of $\beta$.  For example, we show in Fig.~\ref{fig:figure1}(a)-(c) the solutions (\ref{eq:exact}) for $\beta=0.1$, $1$ and $10$ (solid black curves), where the arrow indicates increasing time.  There are qualitatively similar features in each case.  For example, the solutions have the property $\partial u/\partial x=0$ at $x=0$, corresponding to no flux at the left boundary.  Further, the solutions each have $u=0$ at $x=\pi/2\sqrt{1+\beta}$, which is a Dirichlet condition at the right boundary.  Thus, physically speaking, there is a loss of \textrm{mass} at the right boundary and indeed $u$ continues to decay until the population becomes extinct as $t\rightarrow\infty$.  Note the domain is decreasing in size as $\beta$ increases.  Also included in Fig.~\ref{fig:figure1}(a)-(c) are numerical solutions (green dashed), computed using finite differences with a no-flux and Dirichlet conditions at the left and right boundaries, respectively, with the details included on \href{https://github.com/ProfMJSimpson/FisherKPP}{GitHub}.  Clearly there is a very good match, confirming the derivation of the exact solution.  We return to Fig.~\ref{fig:figure1} shortly.

\subsection{Asymptotic limits}\label{sec:limits}

The regime $\beta\gg 1$ is interesting since
$$
D(u)\sim 1-\frac{u}{\beta},\quad R(u)\sim u(1-u)+\frac{u^2(1-u)}{2\beta},
\quad\mbox{as}\quad \beta\rightarrow\infty.
$$
That is, for large $\beta$, to leading order the diffusion term is constant and the reaction term is logistic.  Therefore, the reaction-diffusion equation (\ref{eq:generalreactiondiffusion}) with (\ref{eq:DandR}) is a close approximation of the Fisher-KPP equation (\ref{eq:FisherKPP}).  In this limit, from (\ref{eq:exact}) we have
$u\sim u_0\mathrm{e}^{-\beta t}\cos\left(\sqrt{1+\beta}x\right)$ on $0<x<(\pi/2\sqrt{1+\beta})$ as $\beta\rightarrow\infty$.  Note the domain length continues to shrink as $\beta$ increases.

In the limit $\beta\rightarrow 0^+$, we have $\pi/(2\sqrt{1+\beta})\rightarrow\pi/2^-$, so the domain for the exact solution (\ref{eq:exact}) is roughly $0<x<\pi/2$.  For $\beta\ll 1$, the nonlinear diffusivity $D(u)$ is not close to being constant, as we can see from Fig.~\ref{fig:figure0}(a).  While the solution $u(x,t)$ decays very slowly for $\beta\ll 1$, ultimately for $t\gg 1/\beta$ the population density becomes small and $u\sim \beta\ln\left(1+u_0/\beta\right)\mathrm{e}^{-\beta t}\cos\left(\sqrt{1+\beta}x\right)$ in this regime.

\subsection{Illustrative example}\label{sec:example}

As is common with exact solutions that come from symmetry analysis, we are not free to choose our initial condition.  However, in our case we are free to choose $u_0$ to approximate a practical example.  Suppose we employ the obvious initial condition from a mathematical modelling perspective, namely $u=1$ for $0<x<L$. In this case the initial \textrm{mass} of the solution is $L$. With this in mind, we choose $u_0$ in our exact solution (\ref{eq:exact}) so the initial mass of the solution is $\pi/(2\sqrt{1+\beta})$ (remembering that $L=\pi/(2\sqrt{1+\beta})$ for our solution).  By integrating (\ref{eq:exact}) at $t=0$ from $x=0$ to $0<x<\pi/(2\sqrt{1+\beta})$, we arrive at the nonlinear algebraic equation
\begin{equation}
\dfrac{1}{\beta} + 1 = J_0\left(\textrm{ln}\left(\dfrac{\beta + u_0}{\beta} \right) \right)+ L_0\left(\textrm{ln}\left(\dfrac{\beta + u_0}{\beta} \right) \right),
\label{eq:initialmasssolution}
\end{equation}
where $J_0(x)$ is the Bessel function of the first kind and $L_0(x)$ is a modified Struve function of the first kind.  For a given $\beta$, we can solve (\ref{eq:initialmasssolution}) for $u_0$.  The result of these calculations is illustrated in Fig.~\ref{fig:figure1}(d)-(f).  Here, numerical solutions (green dashed) computed for the physically interesting initial condition are compared to the exact solution with $u_0$ found from (\ref{eq:initialmasssolution}).  While there is no match for small time, we argue that the exact solutions (\ref{eq:exact}) provide reasonably good approximations for the numerical solutions for intermediate to large times.

\section{Discussion}\label{sec:discussion}

In this article we have analysed a one-parameter family of exact solutions (\ref{eq:exact}) of the reaction-diffusion model (\ref{eq:generalreactiondiffusion}), where the nonlinear diffusion and reaction terms are given by (\ref{eq:DandR}).  To put this work into context, all of these exact solutions are qualitatively similar to solutions of the Fisher-KPP model
\begin{equation}
\frac{\partial u}{\partial t}=\frac{\partial^2 u}{\partial x^2}+u(1-u),
\quad 0<x<L,
\label{eq:Fisher1}
\end{equation}
\begin{equation}
\frac{\partial u}{\partial x}=0 \quad\mbox{on}\quad x=0,
\label{eq:Fisher2}
\end{equation}
\begin{equation}
u=0, \quad\mbox{on}\quad x=L,
\label{eq:Fisher3}
\end{equation}
\begin{equation}
u(x,0)=f(x),\quad 0<x<L,
\label{eq:Fisher4}
\end{equation}
provided $L<\pi/2$.  Indeed, for the one-parameter family of exact solutions (\ref{eq:exact})-(\ref{eq:DandR}) and all solutions to the Fisher-KPP model (\ref{eq:Fisher1})-(\ref{eq:Fisher4}), we have
\begin{equation}
u\sim U_1\mathrm{e}^{-(\pi^2/4L^2-1)t}\cos\left(\frac{\pi x}{2L}\right)
\quad\mbox{as}\quad t\rightarrow\infty,
\label{eq:latetime}
\end{equation}
where $L$ is: any constant $0<L<\pi/2$ for the Fisher-KPP model; related to $\beta>0$ for our exact solutions (\ref{eq:exact}) via $L=\pi/(2\sqrt{1+\beta})$.  Further, the constant $U_1$ in (\ref{eq:latetime}) is: related to $f(x)$ in (\ref{eq:Fisher4}) in Fisher-KPP in a complicated way; or given by $U_1=\beta\ln\left(1+u_0/\beta\right)$ for our exact solutions (\ref{eq:exact}).  Therefore, in this sense, the exact solutions (\ref{eq:exact}) are perfectly sensible and consistent with previous understanding of these types of reaction-diffusion models (including the extinction property $L<\pi/2$, for example \cite{Bradford1970,Elhachem2019,Li2022}).

There is the usual downside that come from symmetry analysis, which is that the initial condition of the exact solution is embedded.  We are not free to change the initial condition to suit the physical application or experimental data, for example.  Regardless, the initial condition for (\ref{eq:exact}), namely
\begin{equation}
u(x,0)=\beta\left[\mathrm{exp}\left(\ln\left(1+\frac{u_0}{\beta}\right)\cos \left(\sqrt{1+\beta}\,x\right)\right)-1\right],
\label{eq:exactIC}
\end{equation}
is a decreasing function with properties $u(0,0)=u_0$ and $u(L,0)=0$, which is certainly reasonable, provided $u_0=\mathcal{O}(1)$.  Of course, what makes the exact solutions (\ref{eq:exact}) exceptional and worth recording is not whether they match a particular initial condition, but rather that fully explicit time-dependent solutions to nonlinear reaction-diffusion models are very rare.  Further, apart from explicitly showing how the solution evolves, these formulae can also be used as benchmarks for numerical simulations, for example in courses for numerical methods in nonlinear pdes.

We close by mentioning that the particular nonclassical symmetry used to construct solution \eqref{eq:nonclasssolution} is valid in any number of dimensions and in any coordinate system \cite{Broadbridge2015}, so that solutions to equation \eqref{eq:generalreactiondiffusion} with nonlinear diffusivity and reaction given by \eqref{eq:diffusionreaction} could be constructed in $\mathbb{R}^n$. For example, in radially symmetric coordinates in $\mathbb{R}^2$, solution \eqref{eq:nonclasssolution} will still be valid, however the solution of the Helmholtz equation, $\Psi(r)$, will be written in terms of Bessel functions (in $\mathbb{R}^3$ the solution will be in terms of spherical Bessel functions \cite{BradshawHajek2020}). The higher dimensional analogue of many of the other results presented here can also be produced, for example, the solution will decay (and become extinct) if $L<\lambda/\sqrt{1+\beta}$ where $\lambda$ is the first zero of the Bessel function, $J_0(x)$.\\

\noindent
\textit{Acknowledgements.} This work is supported by the Australian Research Council (DP200100177, DP200102130). \\

\noindent
\textit{Supplementary Data.} Supplementary material is available at \href{https://github.com/ProfMJSimpson/FisherKPP}{GitHub}. \\

\end{document}